\documentclass[12pt]{article}
\usepackage{graphicx}
\usepackage{tikz}

\newcommand\pubnumber{}
\newcommand\pubdate{\today}

\def\mcr{School of Physics and Astronomy\\
The University of Manchester, Manchester, M13 9PL, UK}
\def\support{\footnote{The author is supported by STFC, UK, under grant ST/J004332/1.}}

\def\Title#1{\begin{center} {\Large #1 } \end{center}}
\def\Author#1{\begin{center}{ \sc #1} \end{center}}
\def\Address#1{\begin{center}{ \it #1} \end{center}}

\newcommand\pubblock{\rightline{\begin{tabular}{l} \pubnumber\\
         \pubdate  \end{tabular}}}
\newenvironment{Abstract}{\begin{quotation}  }{\end{quotation}}
\newenvironment{Presented}{\begin{quotation} \begin{center} 
             PRESENTED AT\end{center}\bigskip 
      \begin{center}\begin{large}}{\end{large}\end{center} \end{quotation}}
\def\Acknowledgements{\bigskip  \bigskip \begin{center} \begin{large}
             \bf ACKNOWLEDGEMENTS \end{large}\end{center}}
\usepackage{ifthen} 
\newboolean{pdflatex}
\setboolean{pdflatex}{true} 

\newboolean{articletitles}
\setboolean{articletitles}{true} 

\newboolean{uprightparticles}
\setboolean{uprightparticles}{false} 
\usepackage{microtype}
\usepackage{lineno}  
\usepackage{xspace} 

\usepackage{graphicx}  
\usepackage{color}
\usepackage{colortbl}
\graphicspath{{./figs/}} 

\usepackage{amsmath} 
\usepackage{amssymb}
\usepackage{amsfonts}
\usepackage{upgreek} 

\newcommand*\patchAmsMathEnvironmentForLineno[1]{%
\expandafter\let\csname old#1\expandafter\endcsname\csname #1\endcsname
\expandafter\let\csname oldend#1\expandafter\endcsname\csname
end#1\endcsname
 \renewenvironment{#1}%
   {\linenomath\csname old#1\endcsname}%
   {\csname oldend#1\endcsname\endlinenomath}%
}
\newcommand*\patchBothAmsMathEnvironmentsForLineno[1]{%
  \patchAmsMathEnvironmentForLineno{#1}%
  \patchAmsMathEnvironmentForLineno{#1*}%
}
\AtBeginDocument{%
\patchBothAmsMathEnvironmentsForLineno{equation}%
\patchBothAmsMathEnvironmentsForLineno{align}%
\patchBothAmsMathEnvironmentsForLineno{flalign}%
\patchBothAmsMathEnvironmentsForLineno{alignat}%
\patchBothAmsMathEnvironmentsForLineno{gather}%
\patchBothAmsMathEnvironmentsForLineno{multline}%
}

\usepackage{hyperref}    
\usepackage[all]{hypcap} 

\def\babar  {\mbox{BaBar}\xspace}

\ifthenelse{\boolean{uprightparticles}}%
{

 \def\Ppi         {\ensuremath{\uppi}\xspace}

 \def\PDelta      {\ensuremath{\Delta}\xspace}                 
 \def\PXi      {\ensuremath{\Xi}\xspace}                 
 \def\PLambda      {\ensuremath{\Lambda}\xspace}                 
 \def\PSigma      {\ensuremath{\Sigma}\xspace}                 
 \def\POmega      {\ensuremath{\Omega}\xspace}                 
 \def\PUpsilon      {\ensuremath{\Upsilon}\xspace}                 
 
 \def\PB      {\ensuremath{\mathrm{B}}\xspace}                 
                  
 \def\PD      {\ensuremath{\mathrm{D}}\xspace}

 \def\PK      {\ensuremath{\mathrm{K}}\xspace}

 \def\Pi      {\ensuremath{\mathrm{i}}\xspace}

 \def\Ps      {\ensuremath{\mathrm{s}}\xspace}

}
{

 \def\Ppi         {\ensuremath{\pi}\xspace}

 \mathchardef\PDelta="7101
 \mathchardef\PXi="7104
 \mathchardef\PLambda="7103
 \mathchardef\PSigma="7106
 \mathchardef\POmega="710A
 \mathchardef\PUpsilon="7107
                  
 \def\PB      {\ensuremath{B}\xspace}                 
                  
 \def\PD      {\ensuremath{D}\xspace}

 \def\PK      {\ensuremath{K}\xspace}

 \def\Pi      {\ensuremath{i}\xspace}

 \def\Ps      {\ensuremath{s}\xspace}

}

\def\squark    {\ensuremath{\Ps}\xspace}

\def\pion  {\ensuremath{\Ppi}\xspace}

\def\pip   {\ensuremath{\pion^+}\xspace}
\def\pim   {\ensuremath{\pion^-}\xspace}

\def\kaon  {\ensuremath{\PK}\xspace}
  \def\Kbar  {\kern 0.2em\overline{\kern -0.2em \PK}{}\xspace}

\def\Kp    {\ensuremath{\kaon^+}\xspace}
\def\Km    {\ensuremath{\kaon^-}\xspace}

\def\KS    {\ensuremath{\kaon^0_{\rm\scriptscriptstyle S}}\xspace}

  \def\Dbar    {\kern 0.2em\overline{\kern -0.2em \PD}{}\xspace}
\def\D       {\ensuremath{\PD}\xspace}

\def\Dz      {\ensuremath{\D^0}\xspace}
\def\Dzb     {\ensuremath{\Dbar^0}\xspace}
\def\Dp      {\ensuremath{\D^+}\xspace}

\def\Dsp     {\ensuremath{\D^+_\squark}\xspace}

\def\Bbar    {\ensuremath{\kern 0.18em\overline{\kern -0.18em \PB}{}}\xspace}

  \def\Y#1S{\ensuremath{\PUpsilon{(#1S)}}\xspace}

\def\Lbar {\ensuremath{\kern 0.1em\overline{\kern -0.1em\PLambda}}\xspace}

\newcommand{\decay}[2]{\ensuremath{#1\!\to #2}\xspace}         

\def\to                 {\ensuremath{\rightarrow}\xspace}

\def\CP                {\ensuremath{C\!P}\xspace}

\def\AT#1     {\ensuremath{A_{\mathrm{T}}^{#1}}\xspace}           

\def\C#1      {\ensuremath{\mathcal{C}_{#1}}\xspace}                       
\def\Cp#1     {\ensuremath{\mathcal{C}_{#1}^{'}}\xspace}                    
\def\Ceff#1   {\ensuremath{\mathcal{C}_{#1}^{\mathrm{(eff)}}}\xspace}        
\def\Cpeff#1  {\ensuremath{\mathcal{C}_{#1}^{'\mathrm{(eff)}}}\xspace}       
\def\Ope#1    {\ensuremath{\mathcal{O}_{#1}}\xspace}                       
\def\Opep#1   {\ensuremath{\mathcal{O}_{#1}^{'}}\xspace}                    

\newcommand{\tev}{\ifthenelse{\boolean{inbibliography}}{\ensuremath{~T\kern -0.05em eV}\xspace}{\ensuremath{\mathrm{\,Te\kern -0.1em V}}\xspace}}
\newcommand{\gev}{\ensuremath{\mathrm{\,Ge\kern -0.1em V}}\xspace}
\newcommand{\mev}{\ensuremath{\mathrm{\,Me\kern -0.1em V}}\xspace}
\newcommand{\kev}{\ensuremath{\mathrm{\,ke\kern -0.1em V}}\xspace}
\newcommand{\ev}{\ensuremath{\mathrm{\,e\kern -0.1em V}}\xspace}
\newcommand{\gevc}{\ensuremath{{\mathrm{\,Ge\kern -0.1em V\!/}c}}\xspace}
\newcommand{\mevc}{\ensuremath{{\mathrm{\,Me\kern -0.1em V\!/}c}}\xspace}
\newcommand{\gevcc}{\ensuremath{{\mathrm{\,Ge\kern -0.1em V\!/}c^2}}\xspace}
\newcommand{\gevgevcccc}{\ensuremath{{\mathrm{\,Ge\kern -0.1em V^2\!/}c^4}}\xspace}
\newcommand{\mevcc}{\ensuremath{{\mathrm{\,Me\kern -0.1em V\!/}c^2}}\xspace}

\def\gsim{{~\raise.15em\hbox{$>$}\kern-.85em
          \lower.35em\hbox{$\sim$}~}\xspace}
\def\lsim{{~\raise.15em\hbox{$<$}\kern-.85em
          \lower.35em\hbox{$\sim$}~}\xspace}

\def\tell1  {TELL1\xspace}
\def\ukl1   {UKL1\xspace}

\usepackage{cite} 
\usepackage{mciteplus}

\begin{document}
\begin{titlepage}
\pubblock

\vfill
\Title{HFAG-charm averages}
\vfill
\Author{ Marco Gersabeck\support~on behalf of the HFAG-charm group}
\Address{\mcr}
\vfill
\begin{Abstract}
The extraction of charm mixing and \CP violation parameters requires the combination of many measurements.
The latest averages are reviewed and the contribution of individual measurements is discussed.
While it is established that the \CP-even eigenstate is the shorter lived, the oscillation frequency of charm mesons is still not known to high accuracy.
All \CP asymmetries are found to be compatible with zero within less than $3\sigma$.
\end{Abstract}
\vfill
\begin{Presented}
The \\
8$^{\rm th}$ International Workshop on the CKM Unitarity Triangle \\
(CKM 2014)\\
Vienna, Austria, September 8-12, 2014
\end{Presented}
\vfill
\end{titlepage}
\def\thefootnote{\fnsymbol{footnote}}
\setcounter{footnote}{0}

\section{Introduction}

The charm sub-group of the Heavy Flavour Averaging Group~\cite{Amhis:2012bh} provides averages and overviews of measurements in a diverse range of areas.
These include mixing and \CP violation, which are reviewed in more detail here, but also properties of excited charm states, leptonic and semi-leptonic charm decays, hadronic branching fractions, as well as baryonic, rare, and forbidden decays.
The following section covers the extraction of mixing and indirect \CP violation parameters, followed by a short summary of time-integrated asymmetries.

\section{Mixing and indirect \CP violation}
The mass eigenstates of neutral charm mesons are denoted as $|\PD_1\rangle \equiv p|\Dz\rangle + q|\Dzb\rangle$ and $|\PD_2\rangle \equiv p|\Dz\rangle - q|\Dzb\rangle$, with complex coefficients satisfying $|p|^2+|q|^2=1$.
The parameter $\phi$ is the weak phase difference ${\rm arg}(q/p)$.
The mixing parameters are defined as $x \equiv (m_2 − m_1)/\Gamma$ and $y \equiv (\Gamma_2 − \Gamma_1)/(2\Gamma)$, where $\Gamma = (\Gamma_1 + \Gamma_2)/2$. 
The convention used by HFAG is $(\CP)|\Dz\rangle = −|\Dzb\rangle$ and $(\CP)|\Dzb\rangle = −|\Dz\rangle$.
Thus, in the absence of \CP violation, $|\PD_1\rangle$ is \CP-odd and $|\PD_2\rangle$ is \CP-even.

The extraction of the parameters describing mixing and indirect \CP violation in the charm system requires the combination of many measurements.
In total, 45 measurements are used to extract ten parameters whose results are summarised in Table~\ref{tab:results}.
Due to the interplay of direct and indirect \CP violation, some parameters describing direct \CP violation in Cabibbo-suppressed decays have to be included~\cite{Kagan:2009gb,Gersabeck:2011xj}.
The average~\cite{Amhis:2012bh} yields no indication of \CP violation.
Charm mixing on the other hand is well established with $y$ being confirmed to be positive, indicating that the \CP-even state is the shorter lived.
The precision on $x$ is significantly worse and this parameter is currently not significantly non-zero ($2.4\sigma$).
This can be improved particularly through measurements of \decay{\Dz}{\KS\pim\pip} decays as is discussed in more detail below.

\begin{table}
\renewcommand{\arraystretch}{1.4}
\begin{center}
\caption{\label{tab:results}
Results of the global fit for different assumptions concerning~\CP violation. Reproduced from Ref.~\cite{Amhis:2012bh}}
\vspace*{6pt}
\footnotesize
\begin{tabular}{c|cccc}
\hline
\textbf{Parameter} & \textbf{\boldmath No CPV} & \textbf{\boldmath No direct CPV} 
& \textbf{\boldmath CPV-allowed} & \textbf{\boldmath CPV-allowed} \\
 & & \textbf{\boldmath in DCS decays} & & \textbf{\boldmath 95\% CL Interval} \\
\hline
$\begin{array}{c}
x\ (\%) \\ 
y\ (\%) \\ 
\delta^{}_{K\pi}\ (^\circ) \\ 
R^{}_D\ (\%) \\ 
A^{}_D\ (\%) \\ 
|q/p| \\ 
\phi\ (^\circ) \\
\delta^{}_{K\pi\pi}\ (^\circ)  \\
A^{}_{\pi} (\%) \\
A^{}_K (\%) \\
x^{}_{12}\ (\%) \\ 
y^{}_{12}\ (\%) \\ 
\phi^{}_{12} (^\circ)
\end{array}$ & 
$\begin{array}{c}
0.49\,^{+0.14}_{-0.15} \\
0.62\,\pm 0.08 \\
7.8\,^{+9.6}_{-11.1} \\
0.350\,\pm 0.004 \\
- \\
- \\
- \\
18.7\,^{+23.2}_{-23.7} \\
- \\
- \\
- \\
- \\
- 
\end{array}$ &
$\begin{array}{c}
0.43\,^{+0.14}_{-0.15}\\
0.60\,\,\pm 0.07 \\
4.6\,^{+10.3}_{-12.0}\\
0.349\,\pm 0.004 \\
- \\
1.007\,^{+0.015}_{-0.014}\\
-0.30\,^{+0.58}_{-0.60} \\ 
20.8\,^{+23.9}_{-24.3} \\ 
0.11\,\pm 0.14 \\
-0.13\,\pm 0.13 \\
0.43\,^{+0.14}_{-0.15}\\
0.60\,\pm 0.07 \\
0.9\,^{+1.9}_{-1.7} 
\end{array}$ &
$\begin{array}{c}
0.41\,^{+0.14}_{-0.15}\\
0.63\,\,^{+0.07}_{-0.08}\\
7.3\,^{+9.8}_{-11.5} \\
0.349\,\pm 0.004 \\
-0.71\,^{+0.92}_{-0.95} \\
0.93\,^{+0.09}_{-0.08} \\ 
-8.7\,^{+8.7}_{-9.1} \\ 
23.3\,^{+23.9}_{-24.4} \\
0.14\,\pm 0.15 \\
-0.11\,^{+0.14}_{-0.13} \\
 \\
 \\
 \\
\end{array}$ &
$\begin{array}{c}
\left[ 0.11,\, 0.68\right] \\
\left[ 0.47,\, 0.76\right] \\
\left[ -18.5,\, 25.8\right] \\
\left[ 0.342,\, 0.356\right] \\
\left[ -2.6,\, 1.1\right] \\
\left[ 0.79,\, 1.12\right] \\\
\left[ -26.9,\, 8.6\right] \\
\left[ -24.8,\, 70.2\right] \\
\left[ -0.15,\, 0.42\right] \\
\left[ -0.37,\, 0.15\right] \\
\left[ 0.13,\, 0.69\right] \\
\left[ 0.45,\, 0.75\right] \\
\left[ -3.0,\, 6.1\right] \\
\end{array}$ \\
\hline
\end{tabular}
\end{center}
\end{table}

There is a remarkable difference in sensitivity between the scenario allowing for all types of \CP violation and that using the so-called superweak approximation.
The latter introduces the relation $\tan\phi=(1-|q/p|^2)/(1+|q/p|^2)\times(x/y)$, which reduces the number of parameters by one~\cite{Ciuchini:2007cw,Kagan:2009gb,Amhis:2012bh}.
In addition, the parameters can also be expressed as $x_{12}$, $y_{12}$ and $\phi_{12}$, which are more closely connected the parameters of the mass and width matrices describing charm mixing~\cite{Kagan:2009gb,Amhis:2012bh}.

A number of measurements have been added since the last CKM workshop.
These are mostly measurements of two-body hadronic final states apart from an updated measurement of \decay{\Dz}{\KS\pim\pip} decays by the Belle collaboration~\cite{Peng:2014oda}.
New measurements using \decay{\Dz}{\Kp\pim} decays have been provided by the CLEO~\cite{Asner:2012xb}, LHCb~\cite{Aaij:2012nva,Aaij:2013wda}, CDF~\cite{Aaltonen:2013pja}, and Belle~\cite{Ko:2014qvu} collaborations.
The \babar and Belle collaborations updated their measurements of the observables $y_{\CP}$ and $A_\Gamma$ using their full data samples~\cite{Lees:2012qh,Staric:2012ta}.
The LHCb collaboration has published a measurement of $A_\Gamma$ based on a part of their dataset~\cite{Aaij:2013ria}.
Finally, LHCb has released measurements of the difference in direct \CP asymmetry in two-body singly Cabibbo-suppressed decays~\cite{LHCb:2013dka,Aaij:2013bra,Aaij:2014gsa}.
In the following the impact of these measurements is reviewed.

\begin{figure}[bt]
\centering
\includegraphics[width=0.7\textwidth]{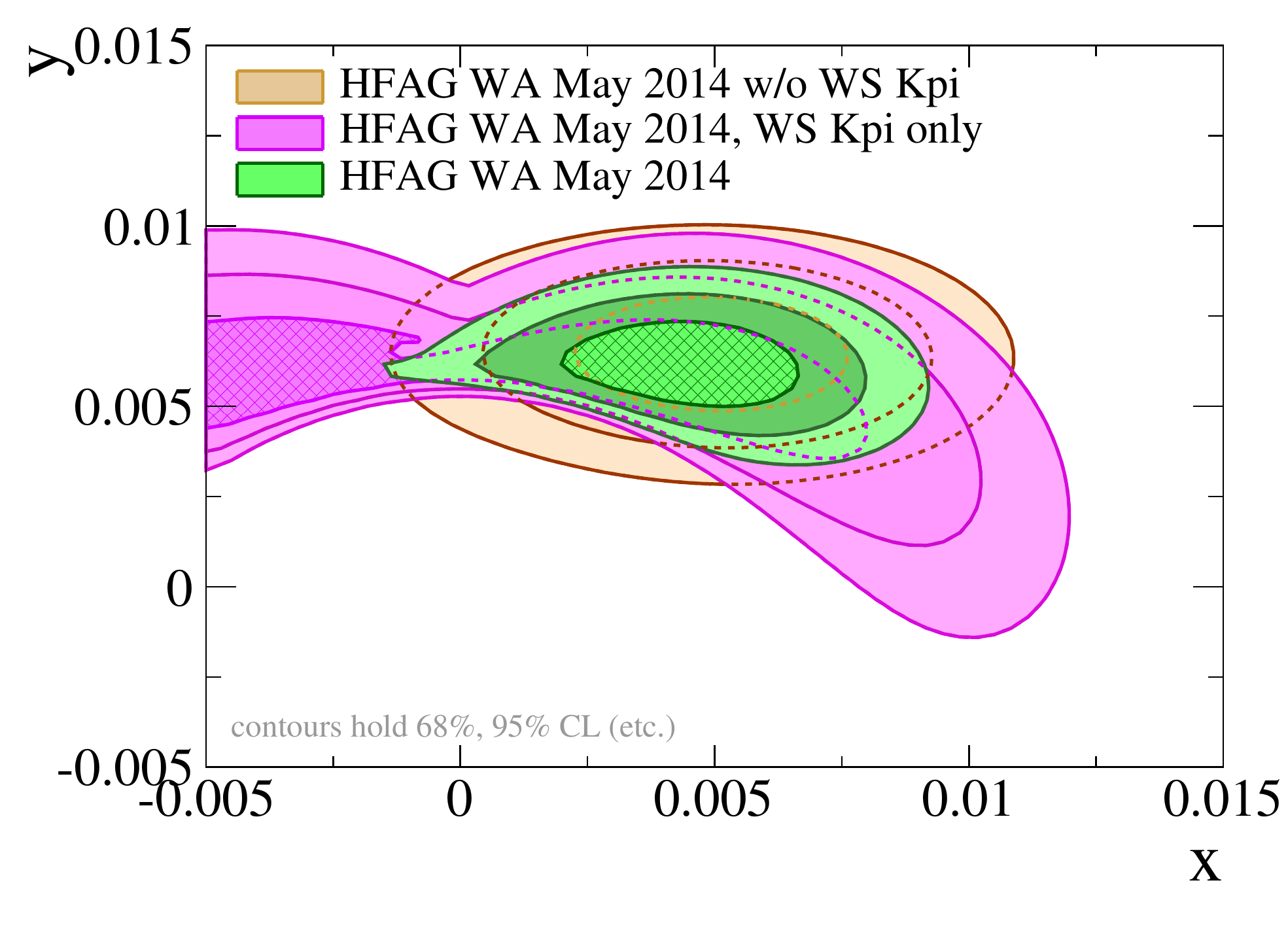}
\caption{Contributions of the combination of different measurements to the extraction of the parameters $x$ and $y$.}
\label{fig:x_y}
\end{figure}

Measurements using \decay{\Dz}{\KS\pim\pip} provide direct access to $x$ and $y$ and thus yield an elliptical contour in a plot showing the allowed ranges for $y$ vs $x$ (see Fig.~\ref{fig:x_y}).
Adding measurements of $y_{\CP}$ constrain $y$ further in the absence of \CP violation.
This is shown as the brown contour in Fig.~\ref{fig:x_y}.
Measurements using \decay{\Dz}{\Kp\pim} decays yield constraints on $y'$ and $x^2+y^2$, where $y'\approx y$ since the strong phase $\delta_{K\pi}$ is known to be close to zero.
This leads to a contour that is symmetrical around $x\approx 0$ and that roughly describes an ellipse around zero for a range in $y$ (see pink contour in Fig.~\ref{fig:x_y}).
The green contour in Fig.~\ref{fig:x_y} shows the full combination of all measurements, to which measurements that were not explicitly mentioned here do not contribute significantly.
This shows that additional measurements of \decay{\Dz}{\KS\pim\pip} decays will be essential to constrain $x$ and to determine its sign with higher significance.

\begin{figure}[tbh]
\centering
\includegraphics[width=0.7\textwidth]{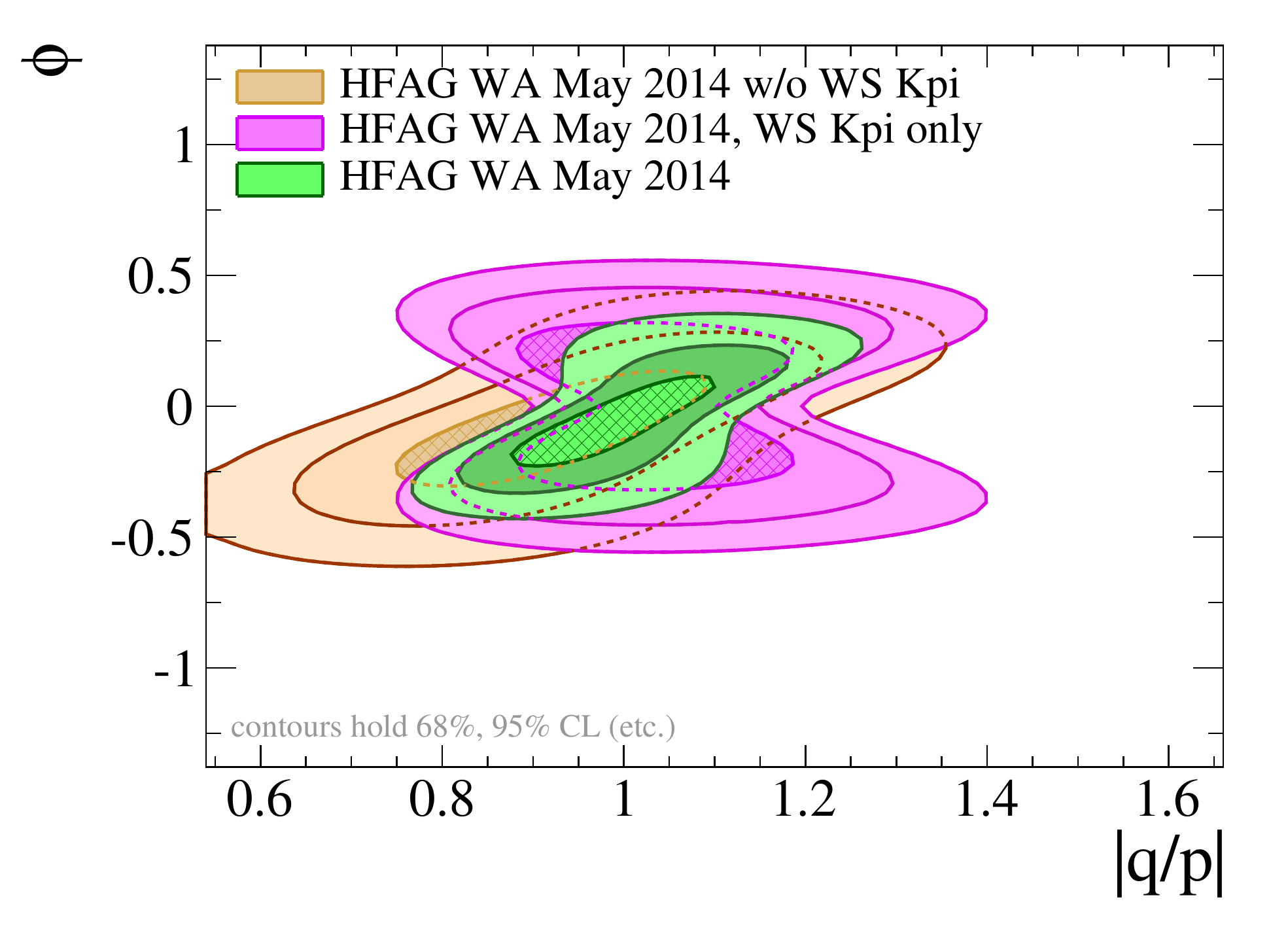}
\caption{Contributions of the combination of different measurements to the extraction of the parameters $|q/p|$ and $\phi$.}
\label{fig:qoverp_phi}
\end{figure}

Similarly to the parameters $x$ and $y$, the \CP violation parameters $|q/p|$ and $\phi$ are measured directly using the decay \decay{\Dz}{\KS\pim\pip}.
The resulting elliptical contour is further constrained in one direction by the addition of measurements of the observable $A_\Gamma$, which determines indirect \CP violation.
This is shown in the brown contour in Fig.~\ref{fig:qoverp_phi}.
Measurements using flavour-tagged \decay{\Dz}{\Kp\pim} decays yield the pink contour in Fig.~\ref{fig:qoverp_phi}, which shows particularly good sensitivity to $|q/p|$ for small absolute values of $\phi$.
The green contour in Fig.~\ref{fig:qoverp_phi} shows the full combination of all measurements, to which measurements that were not explicitly mentioned here do not contribute significantly.

\begin{figure}[tbh]
\centering
\begin{tikzpicture}
\node[anchor=south west,inner sep=0] (image) at (0,0) {\includegraphics[width=0.7\textwidth]{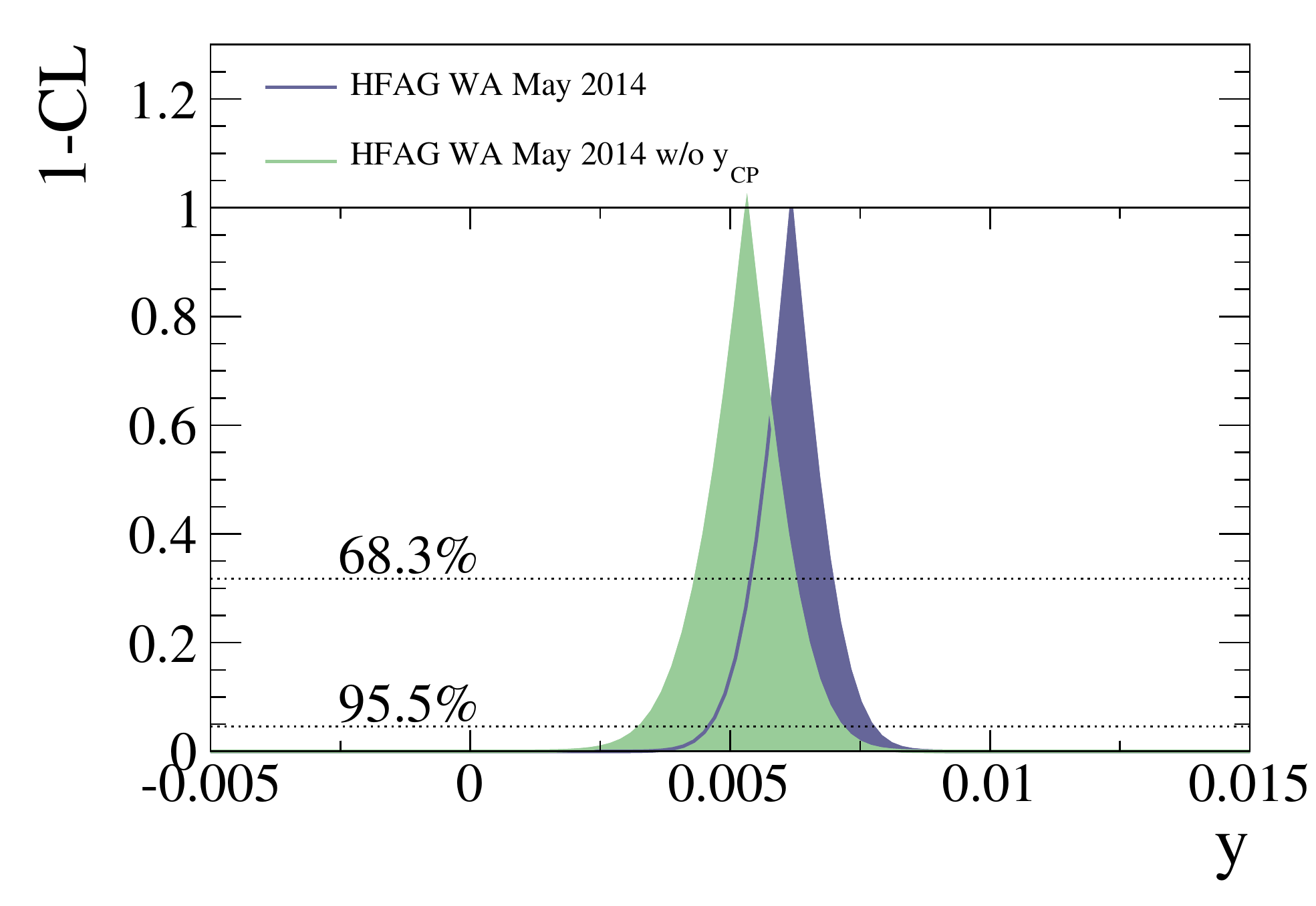}};
 \begin{scope}[x={(image.south east)},y={(image.north west)}]
\draw [thin, draw=black, fill=black, opacity=0.3]
(0.578,0.17) -- (0.578,0.77) -- (0.822,0.77) -- (0.822,0.17) -- cycle;
\draw [thin, draw=black, fill=black, opacity=0.2]
(0.639,0.17) -- (0.639,0.77) -- (0.761,0.77) -- (0.761,0.17) -- cycle;
\node[label, text=white] at (0.70,0.7) {\footnotesize $\pm1\sigma$};
\node[label, text=white] at (0.791,0.7) {\footnotesize $\pm2\sigma$};
\draw [thin, draw=black, fill=black, opacity=0.3]
(0.60,0.88) -- (0.60,0.93) -- (0.70,0.93) -- (0.70,0.88) -- cycle;
\draw [thin, draw=black, fill=black, opacity=0.2]
(0.625,0.88) -- (0.625,0.93) -- (0.675,0.93) -- (0.675,0.88) -- cycle;
\node[label, text=black, anchor=west] at (0.72,0.905) {\scriptsize HFAG $y_{\CP}$};
\node[label, text=black, anchor=west] at (0.72,0.850) {\scriptsize WA May 2014};
\end{scope}
\end{tikzpicture}
\caption{Comparison of the world average for $y$ with and without using measurements of $y_{\CP}$ (shown as likelihood contours) with the world average of $y_{\CP}$ (shown as bars).}
\label{fig:ycp_y}
\end{figure}
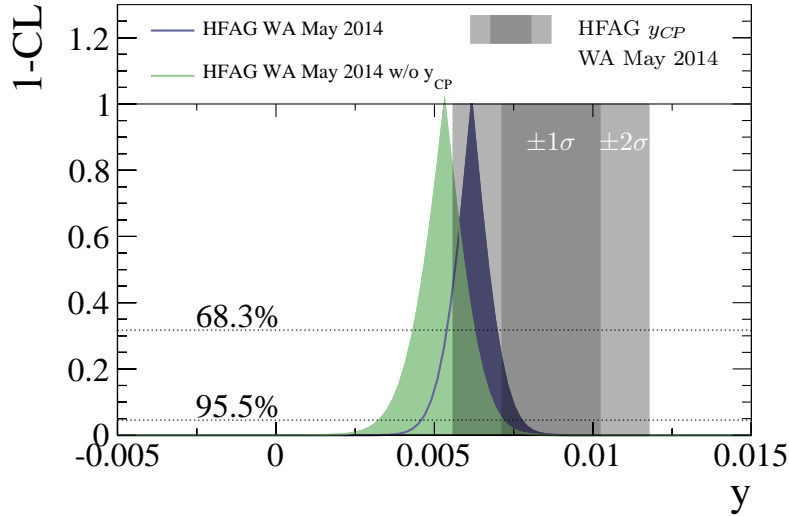

A question arising from these averages is whether there is a tension between the measurements of $y_{\CP}$ and those constraining $y$ directly.
For roughly similar magnitudes of $x$ and $y$ it is expected that the relation $y_{\CP}\lesssim y$ holds unless $|q/p|$ deviates strongly from unity.
As is shown in Fig.~\ref{fig:ycp_y}, the world average of $y$ obtained from all inputs but $y_{\CP}$ does not fully coincide with the world average of $y_{\CP}$.
However, this discrepancy is not very significant ($<2\sigma$).
Thus, while a stronger tension may have existed with older measurements, there is no inconsistency at present.

A dedicated average comparing indirect \CP asymmetry measurements with the difference of direct \CP asymmetries in \Dz decays to $\Km\Kp$ and $\pim\pip$ mesons yields $a_{\CP}^{ind}=(0.013\pm0.052)\%$ and $\Delta a_{\CP}^{dir}=(-0.253\pm0.104)\%$.
With a confidence level for the no \CP violation hypothesis of $5.1\%$, this average also shows no evidence for \CP violation.

\section{Time-integrated asymmetries}
A number of time-integrated asymmetry measurements have been published since the last CKM workshop.
These measurements have been averaged with existing measurements where available.
None of these averages shows a significant deviation from zero.

The range of measurements includes a comprehensive study of hadronic final states of \Dz, \Dp, and \Dsp decays by the CLEO collaboration~\cite{Onyisi:2013bjt,Bonvicini:2013vxi}.
The LHCb collaboration has released measurements of \Dp and \Dsp decays into hadronic final states involving \KS mesons~\cite{Aaij:2014qec}.
The Belle collaboration has published measurements of \Dz decays into two neutral mesons~\cite{Nisar:2014fkc}.

\section{Conclusion}
In summary, there has been enormous progress in improving the precision of charm mixing and \CP violation parameters over the past two years.
However, of the two mixing parameters $x$ remains rather weakly constrained.
Initial evidence for \CP violation in the charm sector is currently not confirmed.
Significant progress is still to be expected from the completion of all of LHCb's measurements based on data taken during the first run of the LHC.
These measurements are likely to dominate the picture at the next CKM workshop.
Beyond that, the LHCb run-2 and upgrade programme as well as Belle II will allow the exploration of \CP asymmetries to a level of precision around $10^{-4}$, thus probing down to the Standard Model level.
In addition, important contributions are expected from BES-III as the only running experiment capable of exploiting quantum-correlated charm-anticharm systems.

\Acknowledgements
The author acknowledges support from Moritz Karbach in preparation of some of the material presented here.

\bibliographystyle{LHCb}
\bibliography{hfag-charm}

\ifx\mcitethebibliography\mciteundefinedmacro
\PackageError{LHCb.bst}{mciteplus.sty has not been loaded}
{This bibstyle requires the use of the mciteplus package.}\fi
\providecommand{\href}[2]{#2}
\begin{mcitethebibliography}{10}
\mciteSetBstSublistMode{n}
\mciteSetBstMaxWidthForm{subitem}{\alph{mcitesubitemcount})}
\mciteSetBstSublistLabelBeginEnd{\mcitemaxwidthsubitemform\space}
{\relax}{\relax}

\bibitem{Amhis:2012bh}
Heavy Flavor Averaging Group, Y.~Amhis {\em et~al.},
  \ifthenelse{\boolean{articletitles}}{{\it {Averages of B-Hadron, C-Hadron,
  and tau-lepton properties as of early 2012}},
  }{}\href{http://arxiv.org/abs/1207.1158}{{\tt arXiv:1207.1158}}\relax
\mciteBstWouldAddEndPuncttrue
\mciteSetBstMidEndSepPunct{\mcitedefaultmidpunct}
{\mcitedefaultendpunct}{\mcitedefaultseppunct}\relax
\EndOfBibitem
\bibitem{Kagan:2009gb}
A.~L. Kagan and M.~D. Sokoloff, \ifthenelse{\boolean{articletitles}}{{\it {On
  indirect \CP violation and implications for \Dz - \Dzb and $\PB_{(s)}$ -
  $\Bbar_{(s)}$ mixing}},
  }{}\href{http://dx.doi.org/10.1103/PhysRevD.80.076008}{Phys.\ Rev.\  {\bf
  D80} (2009) 076008}, \href{http://arxiv.org/abs/0907.3917}{{\tt
  arXiv:0907.3917}}\relax
\mciteBstWouldAddEndPuncttrue
\mciteSetBstMidEndSepPunct{\mcitedefaultmidpunct}
{\mcitedefaultendpunct}{\mcitedefaultseppunct}\relax
\EndOfBibitem
\bibitem{Gersabeck:2011xj}
M.~Gersabeck {\em et~al.}, \ifthenelse{\boolean{articletitles}}{{\it {On the
  interplay of direct and indirect \CP violation in the charm sector}},
  }{}\href{http://dx.doi.org/10.1088/0954-3899/39/4/045005}{J.\ Phys.\  {\bf
  G39} (2012) 045005}, \href{http://arxiv.org/abs/1111.6515}{{\tt
  arXiv:1111.6515}}\relax
\mciteBstWouldAddEndPuncttrue
\mciteSetBstMidEndSepPunct{\mcitedefaultmidpunct}
{\mcitedefaultendpunct}{\mcitedefaultseppunct}\relax
\EndOfBibitem
\bibitem{Ciuchini:2007cw}
M.~Ciuchini {\em et~al.}, \ifthenelse{\boolean{articletitles}}{{\it
  {$\PD-\Dbar$ mixing and new physics: General considerations and constraints
  on the MSSM}},
  }{}\href{http://dx.doi.org/10.1016/j.physletb.2007.08.055}{Phys.\ Lett.\
  {\bf B655} (2007) 162}, \href{http://arxiv.org/abs/hep-ph/0703204}{{\tt
  arXiv:hep-ph/0703204}}\relax
\mciteBstWouldAddEndPuncttrue
\mciteSetBstMidEndSepPunct{\mcitedefaultmidpunct}
{\mcitedefaultendpunct}{\mcitedefaultseppunct}\relax
\EndOfBibitem
\bibitem{Peng:2014oda}
Belle Collaboration, T.~Peng {\em et~al.},
  \ifthenelse{\boolean{articletitles}}{{\it {Measurement of \Dz-\Dzb mixing and
  search for indirect \CP violation using \decay{\Dz}{\KS\pip\pim} decays}},
  }{}\href{http://dx.doi.org/10.1103/PhysRevD.89.091103}{Phys.\ Rev.\  {\bf
  D89} (2014) 091103}, \href{http://arxiv.org/abs/1404.2412}{{\tt
  arXiv:1404.2412}}\relax
\mciteBstWouldAddEndPuncttrue
\mciteSetBstMidEndSepPunct{\mcitedefaultmidpunct}
{\mcitedefaultendpunct}{\mcitedefaultseppunct}\relax
\EndOfBibitem
\bibitem{Asner:2012xb}
CLEO Collaboration, D.~Asner {\em et~al.},
  \ifthenelse{\boolean{articletitles}}{{\it {Updated measurement of the strong
  phase in \decay{\Dz}{\Kp\pim} decay using quantum correlations in
  \decay{e^+e^-}{\Dz\Dzb} at CLEO}},
  }{}\href{http://dx.doi.org/10.1103/PhysRevD.86.112001}{Phys.\ Rev.\  {\bf
  D86} (2012) 112001}, \href{http://arxiv.org/abs/1210.0939}{{\tt
  arXiv:1210.0939}}\relax
\mciteBstWouldAddEndPuncttrue
\mciteSetBstMidEndSepPunct{\mcitedefaultmidpunct}
{\mcitedefaultendpunct}{\mcitedefaultseppunct}\relax
\EndOfBibitem
\bibitem{Aaij:2012nva}
LHCb Collaboration, R.~Aaij {\em et~al.},
  \ifthenelse{\boolean{articletitles}}{{\it {Observation of \Dz-\Dzb
  oscillations}},
  }{}\href{http://dx.doi.org/10.1103/PhysRevLett.110.101802}{Phys.\ Rev.\
  Lett.\  {\bf 110} (2013), no.~10 101802},
  \href{http://arxiv.org/abs/1211.1230}{{\tt arXiv:1211.1230}}\relax
\mciteBstWouldAddEndPuncttrue
\mciteSetBstMidEndSepPunct{\mcitedefaultmidpunct}
{\mcitedefaultendpunct}{\mcitedefaultseppunct}\relax
\EndOfBibitem
\bibitem{Aaij:2013wda}
LHCb collaboration, R.~Aaij {\em et~al.},
  \ifthenelse{\boolean{articletitles}}{{\it {Measurement of \Dz-\Dzb mixing
  parameters and search for \CP violation using \decay{\Dz}{\Kp\pim} decays}},
  }{}\href{http://dx.doi.org/10.1103/PhysRevLett.111.251801}{Phys.\ Rev.\
  Lett.\  {\bf 111} (2013) 251801}, \href{http://arxiv.org/abs/1309.6534}{{\tt
  arXiv:1309.6534}}\relax
\mciteBstWouldAddEndPuncttrue
\mciteSetBstMidEndSepPunct{\mcitedefaultmidpunct}
{\mcitedefaultendpunct}{\mcitedefaultseppunct}\relax
\EndOfBibitem
\bibitem{Aaltonen:2013pja}
CDF Collaboration, T.~A. Aaltonen {\em et~al.},
  \ifthenelse{\boolean{articletitles}}{{\it {Observation of \Dz-\Dzb mixing
  using the CDF II detector}},
  }{}\href{http://dx.doi.org/10.1103/PhysRevLett.111.231802}{Phys.\ Rev.\
  Lett.\  {\bf 111} (2013), no.~23 231802},
  \href{http://arxiv.org/abs/1309.4078}{{\tt arXiv:1309.4078}}\relax
\mciteBstWouldAddEndPuncttrue
\mciteSetBstMidEndSepPunct{\mcitedefaultmidpunct}
{\mcitedefaultendpunct}{\mcitedefaultseppunct}\relax
\EndOfBibitem
\bibitem{Ko:2014qvu}
Belle Collaboration, B.~Ko {\em et~al.},
  \ifthenelse{\boolean{articletitles}}{{\it {Observation of \Dz-\Dzb mixing in
  $e^+e^-$ collisions}},
  }{}\href{http://dx.doi.org/10.1103/PhysRevLett.112.111801}{Phys.\ Rev.\
  Lett.\  {\bf 112} (2014), no.~11 111801},
  \href{http://arxiv.org/abs/1401.3402}{{\tt arXiv:1401.3402}}\relax
\mciteBstWouldAddEndPuncttrue
\mciteSetBstMidEndSepPunct{\mcitedefaultmidpunct}
{\mcitedefaultendpunct}{\mcitedefaultseppunct}\relax
\EndOfBibitem
\bibitem{Lees:2012qh}
BaBar Collaboration, J.~Lees {\em et~al.},
  \ifthenelse{\boolean{articletitles}}{{\it {Measurement of \Dz-\Dzb mixing and
  \CP violation in two-body \Dz decays}},
  }{}\href{http://dx.doi.org/10.1103/PhysRevD.87.012004}{Phys.\ Rev.\  {\bf
  D87} (2013) 012004}, \href{http://arxiv.org/abs/1209.3896}{{\tt
  arXiv:1209.3896}}\relax
\mciteBstWouldAddEndPuncttrue
\mciteSetBstMidEndSepPunct{\mcitedefaultmidpunct}
{\mcitedefaultendpunct}{\mcitedefaultseppunct}\relax
\EndOfBibitem
\bibitem{Staric:2012ta}
Belle Collaboration, M.~Staric, \ifthenelse{\boolean{articletitles}}{{\it {New
  Belle results on \Dz-\Dzb mixing}},
  }{}\href{http://arxiv.org/abs/1212.3478}{{\tt arXiv:1212.3478}}\relax
\mciteBstWouldAddEndPuncttrue
\mciteSetBstMidEndSepPunct{\mcitedefaultmidpunct}
{\mcitedefaultendpunct}{\mcitedefaultseppunct}\relax
\EndOfBibitem
\bibitem{Aaij:2013ria}
LHCb collaboration, R.~Aaij {\em et~al.},
  \ifthenelse{\boolean{articletitles}}{{\it {Measurements of indirect \CP
  asymmetries in \decay{\Dz}{\Km\Kp} and \decay{\Dz}{\pim\pip} decays}},
  }{}\href{http://dx.doi.org/10.1103/PhysRevLett.112.041801}{Phys.\ Rev.\
  Lett.\  {\bf 112} (2014), no.~4 041801},
  \href{http://arxiv.org/abs/1310.7201}{{\tt arXiv:1310.7201}}\relax
\mciteBstWouldAddEndPuncttrue
\mciteSetBstMidEndSepPunct{\mcitedefaultmidpunct}
{\mcitedefaultendpunct}{\mcitedefaultseppunct}\relax
\EndOfBibitem
\bibitem{LHCb:2013dka}
LHCb Collaboration, \ifthenelse{\boolean{articletitles}}{{\it {A search for
  time-integrated \CP violation in \decay{\Dz}{\Km\Kp} and
  \decay{\Dz}{\pim\pip} decays}}, }{}\relax
\mciteBstWouldAddEndPuncttrue
\mciteSetBstMidEndSepPunct{\mcitedefaultmidpunct}
{\mcitedefaultendpunct}{\mcitedefaultseppunct}\relax
\EndOfBibitem
\bibitem{Aaij:2013bra}
LHCb collaboration, R.~Aaij {\em et~al.},
  \ifthenelse{\boolean{articletitles}}{{\it {Search for direct \CP violation in
  \decay{\Dz}{h^-h^+} modes using semileptonic \PB decays}},
  }{}\href{http://dx.doi.org/10.1016/j.physletb.2013.04.061}{Phys.\ Lett.\
  {\bf B723} (2013) 33}, \href{http://arxiv.org/abs/1303.2614}{{\tt
  arXiv:1303.2614}}\relax
\mciteBstWouldAddEndPuncttrue
\mciteSetBstMidEndSepPunct{\mcitedefaultmidpunct}
{\mcitedefaultendpunct}{\mcitedefaultseppunct}\relax
\EndOfBibitem
\bibitem{Aaij:2014gsa}
LHCb collaboration, R.~Aaij {\em et~al.},
  \ifthenelse{\boolean{articletitles}}{{\it {Measurement of \CP asymmetry in
  \decay{\Dz}{\Km\Kp} and \decay{\Dz}{\pim\pip} decays}},
  }{}\href{http://dx.doi.org/10.1007/JHEP07(2014)041}{JHEP {\bf 1407} (2014)
  041}, \href{http://arxiv.org/abs/1405.2797}{{\tt arXiv:1405.2797}}\relax
\mciteBstWouldAddEndPuncttrue
\mciteSetBstMidEndSepPunct{\mcitedefaultmidpunct}
{\mcitedefaultendpunct}{\mcitedefaultseppunct}\relax
\EndOfBibitem
\bibitem{Onyisi:2013bjt}
CLEO Collaboration, P.~Onyisi {\em et~al.},
  \ifthenelse{\boolean{articletitles}}{{\it {Improved measurement of absolute
  hadronic branching fractions of the \Dsp meson}},
  }{}\href{http://dx.doi.org/10.1103/PhysRevD.88.032009}{Phys.\ Rev.\  {\bf
  D88} (2013), no.~3 032009}, \href{http://arxiv.org/abs/1306.5363}{{\tt
  arXiv:1306.5363}}\relax
\mciteBstWouldAddEndPuncttrue
\mciteSetBstMidEndSepPunct{\mcitedefaultmidpunct}
{\mcitedefaultendpunct}{\mcitedefaultseppunct}\relax
\EndOfBibitem
\bibitem{Bonvicini:2013vxi}
CLEO Collaboration, G.~Bonvicini {\em et~al.},
  \ifthenelse{\boolean{articletitles}}{{\it {Updated measurements of absolute
  \Dp and \Dz hadronic branching fractions and $\sigma(e^+e^-\to
  D\overline{D})$ at $E_\mathrm{cm} = 3774$ MeV}},
  }{}\href{http://dx.doi.org/10.1103/PhysRevD.89.072002}{Phys.\ Rev.\  {\bf
  D89} (2014), no.~7 072002}, \href{http://arxiv.org/abs/1312.6775}{{\tt
  arXiv:1312.6775}}\relax
\mciteBstWouldAddEndPuncttrue
\mciteSetBstMidEndSepPunct{\mcitedefaultmidpunct}
{\mcitedefaultendpunct}{\mcitedefaultseppunct}\relax
\EndOfBibitem
\bibitem{Aaij:2014qec}
LHCb Collaboration, R.~Aaij {\em et~al.},
  \ifthenelse{\boolean{articletitles}}{{\it {Search for \CP violation in
  \decay{\Dpm}{\KS\Kpm} and \decay{\Dpm}{\KS\pipm} decays}},
  }{}\href{http://dx.doi.org/10.1007/JHEP10(2014)025}{JHEP {\bf 1410} (2014)
  25}, \href{http://arxiv.org/abs/1406.2624}{{\tt arXiv:1406.2624}}\relax
\mciteBstWouldAddEndPuncttrue
\mciteSetBstMidEndSepPunct{\mcitedefaultmidpunct}
{\mcitedefaultendpunct}{\mcitedefaultseppunct}\relax
\EndOfBibitem
\bibitem{Nisar:2014fkc}
Belle Collaboration, N.~Nisar {\em et~al.},
  \ifthenelse{\boolean{articletitles}}{{\it {Search for \CP violation in
  \decay{\Dz}{\piz\piz} decays}},
  }{}\href{http://dx.doi.org/10.1103/PhysRevLett.112.211601}{Phys.\ Rev.\
  Lett.\  {\bf 112} (2014) 211601}, \href{http://arxiv.org/abs/1404.1266}{{\tt
  arXiv:1404.1266}}\relax
\mciteBstWouldAddEndPuncttrue
\mciteSetBstMidEndSepPunct{\mcitedefaultmidpunct}
{\mcitedefaultendpunct}{\mcitedefaultseppunct}\relax
\EndOfBibitem
\end{mcitethebibliography}
\end{document}